\documentclass[12pt,showpacs,a4paper,prb]{revtex4-1}
\usepackage{amsmath}
\usepackage{amssymb}
\begin{document}
\title{Charge quantization emerging from irreducible representations of the gauge group U(1) and local gauge invariance in 2-spinor lenguage}
\author{J.Buitrago}
\affiliation{ Department of Astrophysics of the University of La Laguna, Avenida
Francisco Sanchez, s/n, 38205, La Laguna, Tenerife, Spain
and Instituto de Astrofisica de Canarias (IAC), E-38200 La Laguna, Tenerife, Spain}

\email{jgb@iac.es} 
\date{\today}
\begin{abstract}

A new classical 2-spinor approach to $U(1)$ gauge theory is presented in which the usual four-potential vector field is replaced by a symmetric second rank spinor. Following a lagrangian formulation, it is shown that the four-rank spinor representing the Maxwell field tensor has a $U(1)$ local gauge invariance in terms of the electric and magnetic field strengths. When applied to the magnetic field of a monopole, this formulation, via the irreducible representations condition for the gauge group, leads to a quantization condition differing by a factor 2 of the one predicted by Dirac  without relying on any kind of singular vector potentials.

\end{abstract}
\maketitle

\section{INTRODUCTION}
In the  usual Lagrangian  approach to gauge theories, the field-particle interaction term involve the introduction of the field potentials associated to the group  symmetry of the field (for instance, the electromagnetic four-potential for $U(1)$ or the potentials, equal in number to the group generators, for the non-abelian groups required in $SU(2)$ or  $SU(3)$). Much have being written and discussed about the role played by the potentials as convenient physical tools for describing physical processes in spite of not being, at least classically, directly related to observation. It is not my intention to add any further comment on this well known issue as one of the main motivations for this work is to present a Weyl 2-spinor approach in which a local $U(1)$ gauge invariance is ultimately related to the electromagnetic $\vec E$ and $\vec B$ fields, without any intervening use of the potentials, via the four-rank $F^{{ABA'B'}}$ hermitian spinor corresponding, in spinor language, to the usual $F^{{\alpha \beta}}$ Maxwell tensor. The local $U(1)$ gauge invariance, introduced in section III, have as starting point a set of coupled spinor equations obtained in a previous work (cited below) which being classical in origin are susceptible of describing intrinsic electronic spin. The mentioned spinor equations, invariant under $U(1)$ local phase transformations, together with the corresponding transformation rule for the symmetric field spinor $\phi^{AB}$, will be used in section IV to approach, in a new way, the old issue of charge quantization (The 2-spinor formalism and notation used here  follows those adopted by Penrose and Rindler \cite{penrose}. Capital index letters take the values 0,1. Indices are raised and lowered with the skew metric spinor $\epsilon^{AB}$ following the rules $\xi^A=\epsilon^{AB}\xi_B$, $\xi_B=\epsilon_{AB}\xi^A$. For any spinor $\xi^A\xi_A=0$).
In order to understand this paper, the reader is asumed to have some familiarity with the relatively unfamiliar 2-spinor formalism and calculus. Some years ago Penrose pointed out (in the book cited above) that 2-spinor calculus apply to a deeper level of structure of space-time than the familiar world-tensor calculus. Now we will see how the 2-spinor lenguage, via the spinor equations, unveils a $U(1)$ local gauge invariance related to the electromagnetic field strengths quantities instead of the potentials. Possibly, in the usual vector-tensor lenguage, this invariance would have never be found.


%
%

%
%
%
\section{SPINOR EQUATIONS}
In a previous article (Ref. 2), the following linear first order differential spinor equations (the derivative is taken respect to the proper time $\tau$) were introduced:
\begin{equation}\label{eqofmot}
\begin{array}{c}
\frac{d\eta^A}{d\tau}=\frac{e}{m}\phi^{AB}\eta_{B} \\
\frac{d\pi^A}{d\tau}=\frac{e}{m}\phi^{AB}\pi_{B}. \\
\end{array}
\end{equation} \\
As shown in Ref. 2, these coupled equations, in natural units $\hbar=c=1$,  are equivalent to the Lorentz Force. However, in the spinor version, describe the motion of a 1/2 spin particle (As it is essential for the rest of this work, this point will be treated in full detail in this section) of mass $m$ and charge $e$ (tipically an electron) under an electromagnetic field described by the symmetric second-rank spinor $\phi^{AB}$,  explicitely given by
\begin{equation} \label{phi}
\phi^{AB}= \frac{1}{2}
\left( \begin{array}{cc}
-\left[ E_{1}+B_{2}\right] + i\left[E_{2}-B_{1}
\right] & E_{3}+iB_{3} \\
E_{3}+iB_{3} & \left[E_{1}-B_{2}\right]+i
\left[B_{1}+E_{2}\right] \\
\end{array}\right) .
\end{equation}
In turn, $\phi^{{AB}}$ and its complex conjugate $\bar\phi^{A^{'}B^{'}}$ form the antisymmetric four-rank electromagnetic field spinor
\begin{equation} \label{emfield}
F^{ABA'B'}=\epsilon^{AB}\bar\phi^{A'B'}+
\epsilon^{A'B'}\phi^{AB},
\end{equation} \\
where $\epsilon^{AB}$ is the spinor metric
\begin{equation}\label{metric}
\epsilon^{AB}=\left(\begin{array}{cc} 0 & 1 \\ -1& 0 \end{array}\right).
\end{equation} \\
The solution of equations (\ref{eqofmot}) determine the 
four momentum of the particle given by the hermitian spinor defined  as superposition of the two null directions $\pi^A\bar\pi^{A'}$ and $\eta^A\bar\eta^{A'}$ as \cite{bette}
\begin{equation} \label{momentum}
p^{AA'}=\frac{1}{\sqrt{2}}\left[\pi^{A}\bar\pi^{A'}+\eta^{A}\bar\eta^{A'}\right].
\end{equation} \\
Since $p^{AA'}$ is to represent the four-momentum of a massive particle, must be time-like and certainly fulfill the condition:
\begin{equation} \label{condition}
p^{AA'}p_{AA'}=m^2.
\end{equation} \\
On the other hand, following the standard representation, the different components of $p^{AA'}$ are labeled according to
\begin{equation}
p^{AA'}=\left( \begin{array}{cc}
p^{00'} & p^{01'}\\
p^{10'} & p^{11'} \\
\end{array}\right)
= \frac{1}{\sqrt 2}\left( \begin{array}{cc}
p^0+p^3 & p^1+ip^2 \\
p^1-ip^2 & p^0-p^3 \\
\end{array}\right).
\end{equation}
These last expressions must be used when solving an specific case, via the spinor equations, to identify the components of $p^{AA'}$ in the solution. 

Now, taking into account the precedent relations and irrespective of any specific representation of $p^{AA'}$, the spinor translation of the Lorentz Force equation in tensor form
\begin{equation}
\frac {d}{d\tau }p^{\alpha }=\frac{e}{m}F^{\alpha \beta }p_{\beta },
\end{equation}\\
is given by
\begin{equation}
\dot p^{AA'}=\frac{e}{m}F^{ABA'B'}p_{BB'},
\end{equation}
(the dot meaning derivative respect to proper time). In any case, if the equations (\ref{eqofmot}) are to be equivalent to the Lorentz Force equation they must lead to the last expression, which is just the spinor transcription of the familiar tensor equation. This is clearly seen, making use of (\ref{eqofmot}), in the following quick calculation in which, for short, I take $e/m=1$ and drop the $ 1/{\sqrt{2}}$ factor:
\begin{equation} \label{1.8}
\begin{array}{c}
\dot{p}^{AA'}=\dot{\pi}^{A}\bar{\pi}^{A'}+\pi^{A}\dot{\bar{\pi}}^{A'}+
\dot{\eta}^{A}\bar{\eta}^{A'}+\eta^{A}\dot{\bar{\eta}}^{A'} \\
=\phi^{AB}\pi_{B}\bar{\pi}^{A'}+\bar{\phi}^{A'B'}\bar{\pi}_{B'}\pi^{A}+
\phi^{AB}\eta_{B}\bar{\eta}^{A'}+\bar{\phi}^{A'B'}\bar{\eta}_{B'}\eta^{A} \\
=\epsilon^{A'B'}\phi^{AB}\pi_{B}\bar{\pi}_{B'}+
\epsilon^{AB}\bar{\phi}^{A'B'}\pi_{B}\bar{\pi}_{B'} +
\epsilon^{A'B'}\phi^{AB}\eta_{B}\bar{\eta}_{B'} +
\epsilon^{AB}\bar{\phi}^{A'B'}\eta_{B}\bar{\eta}_{B'} \\
=\left( \epsilon^{AB}\bar{\phi}^{A'B'}  + \epsilon^{A'B'}\phi^{AB} \right)
p_{BB'} \\
=F^{ABA'B'}p_{BB'}.
\end{array}
\end{equation}\\
There is an important difference between the tensor form of the Lorentz Force and the spinor equations. In the first case, it does not make any sense to ask about the behaviour of a particle {\it at rest} under the action of a magnetic field. However, as we shall see, equations (\ref{eqofmot}) give a solution describing spin precession. This is a consequence of the fact that any tensor equation can be translated into the 2-spinor form while the opposite is not in general true \cite{carmeli}. In fact, by simple inspection of the 2-spinor equations of motion (\ref{eqofmot}), it is evident that they do not have any tensor counterpart.

For a constant magnetic field $\vec B$ in a {\it fixed} direction given by the unit vector $\widehat {n}=\left( \sin \theta \cos \varphi ,\sin \theta \sin \varphi ,\cos \theta \right)$, the field spinor $\phi^{AB}$, in spherical coordinates $(r,\theta,\varphi)$, from (\ref{phi}), can be found to be given by

\begin{equation}\label{monopole}
\phi^{AB} = i\frac{1}{2}B{\left(\begin{array}{cc} -\sin\theta e^{-i\varphi}& \cos\theta \\ \cos\theta & \sin\theta e^{i\varphi}\end{array}\right)}.
\end{equation} \\
If we lower the second spinor index and multiply by {\it i}, we get
\begin{equation}\label{monopole2}
{\tilde\phi}^A_{\ B}\equiv -i\phi^A_{\ B} = -\frac{1}{2}B{\left(\begin{array}{cc} \cos\theta &\sin\theta e^{-i\varphi} \\  \sin\theta e^{i\varphi} & -\cos\theta \end{array}\right)}.
\end{equation} \\
The first of equations (\ref{eqofmot}) written in the {\it physical} components corresponding to contravariant spinors can now be written as
\begin{equation}\label{eqmot2}
i\dot\eta^A=-\frac{e}{m}{\tilde\phi}^A_{\ B}\eta^B.
\end{equation}\\
In ${\tilde\phi}^A_{\ B}$ we recognize the familiar spin operator in a direction given by the unit vector $\widehat {n}$. The reason behind this, as outlined in Ref. 2, is that in the usual tensor lenguage, the magnetic field vector is related to rotations of the spatial part of the four-momentum $p^\alpha$, therefore to the $SO(3)$ rotation subgroup of the Lorentz group. In Spinor Space $S(2,C)$, $\vec B$ is related (for a particle at rest) to internal rotations of the spinors, hence to $SU(2)$ as universal covering group of $SO(3)$. In the general case of electric and magnetic fields, the relation is between the whole Lorentz Group and $SL(2,C)$.

If $\theta=0$, there are two kind of solutions depending on whether the particle is at rest or in motion. In the first case we have
\begin{equation}
p^{AA'}=\left( \begin{array}{cc}
p^{00'} & p^{01'}\\
p^{10'} & p^{11'} \\
\end{array}\right)
= \frac{1}{\sqrt 2}\left( \begin{array}{cc}
m & 0\\
0 & m \\
\end{array}\right).
\end{equation}
Together with
\begin{equation}\label{monopole2}
{\tilde\phi}^A_{\ B}\equiv -i\phi^A_{\ B} = -\frac{1}{2}B{\left(\begin{array}{cc} 1&0 \\ 0 & -1 \end{array}\right)}.
\end{equation} \\
The solutions of (\ref{eqmot2}) and a similar equation for $\pi^A$, taking into account (\ref{momentum}) and the condition (\ref{condition}) are
\begin{equation}
\pi^{A}(\tau)=\sqrt{m}\ exp \left({-i\frac{e}{2m}B\tau}\right)\left(\begin{array}{c}
1 \\
0 \\
\end{array}\right),\quad \eta^{A}(\tau)=\sqrt{m}\ exp \left({i\frac{e}{2m}B\tau}\right)\left(\begin{array}{c}
0 \\
1 \\
\end{array}\right),
\end{equation} \\
being eigenspinors of the spin operator $S_3$ with eigenvalues $\pm 1/2$ respectively. For the frequency we have
\begin{equation}
\omega=\frac{e}{m}B.
\end{equation}
This frequency is the corresponding to spin precession of a spin one-half electron with a g-factor 2 (the anomalous magnetic moment can only be accounted for in QED). The 1/2 factor in the solutions is due to the peculiar topology of fermions which need a $4\pi$ rotation to return to their original state 
(the same result can eventually be obtained with the Dirac equation although the calculations are more involved).

Next we consider the case $\vec p \not= 0$. The solutions are now:
\begin{displaymath}
\begin{array}{c}
\pi^{0}(\tau)=\pi^{0}(\tau=0)exp \left(
-i \frac{\mbox{\normalsize{$eB$}}}{\mbox{\normalsize{$2m$}}} \tau \right),
{ } \\
{ } \\
\pi^{1}(\tau)=\pi^{1}(\tau=0)exp \left(
i \frac{\mbox{\normalsize{$eB$}}}{\mbox{\normalsize{$2m$}}} \tau \right),
{ } \\
{ } \\
\eta^{0}(\tau)=\eta^{0}(\tau=0)exp \left(
-i \frac{\mbox{\normalsize{$eB$}}}{\mbox{\normalsize{$2m$}}} \tau \right),
{ } \\
{ } \\
\eta^{1}(\tau)=\eta^{1}(\tau=0)exp \left(
i \frac{\mbox{\normalsize{$eB$}}}{\mbox{\normalsize{$2m$}}} \tau \right),
\end{array}
\end{displaymath}
and therefore \\
\begin{displaymath}
\begin{array}{c}
p^{00'}=\vert \pi^{0}(\tau=0)\vert^{2}+ \vert \eta^{0}(\tau=0)\vert^{2}=
constant, \\
{ } \\
p^{01'}=\pi^{0}(\tau=0)\bar{\pi}^{1'}(\tau=0) exp \left(-i\frac{eB}{m}\tau\right)+
\eta^{0}(\tau=0)\bar{\eta}^{1'}(\tau=0) exp \left(-i\frac{eB}{m}\tau\right)=
A(\tau=0)exp \left(-i\frac{eB}{m}\tau \right), \\
{ } \\
p^{11'}=\vert \pi^{1}(\tau=0)\vert^{2}+ \vert \eta^{1}(\tau=0)\vert^{2}=
constant, \\
\end{array}
\end{displaymath}
{ } \\
where $A(\tau=0)$ is a complex number whose value is to be determined from
the initial conditions. The $p^{01'}$ component can thus be written in the form
{ } \\
\begin{equation} \label{4.4}
p^{01'}(\tau)=A_{1}\cos\left(\frac{eB}{m}\tau\right)+A_{2}\sin\left(\frac{eB}{m}\tau\right)+
i\left[ A_{2}\cos\left(\frac{eB}{m}\tau\right)-A_{1}\sin\left(\frac{eB}{m}\tau\right)\right].
\end{equation}
{ } \\
With $A_{1}=Re\{A(\tau=0)\}$ and $A_{2}=Im\{A(\tau=0)\}$. We get finally
for the components of the four-momentum:
\begin{equation} \label{4.5}
\left\{ \begin{array}{c}
p^{00'}= \frac{1}{\sqrt{2}}(E+p_{3})=constant \\
p^{11'}= \frac{1}{\sqrt{2}}(E-p_{3})=constant \\
\end{array}\right\} \Longrightarrow
E,p_{3}=constant,
\end{equation}
{ } \\
\begin{equation} \label{4.6}
p^{01'}=\frac{1}{\sqrt{2}}(p_{1}+ip_{2}) \Longrightarrow
\left\{ \begin{array}{c}
p_{1}= \left[A_{1}\cos(\frac{eB}{m}\tau)+A_{2}\sin(\frac{eB}{m}\tau) \right]
{ } \\
p_{2}=\left[A_{2}\cos(\frac{eB}{m}\tau)-A_{1}\sin(\frac{eB}{m}\tau)\right]
\end{array} \right\},
\end{equation}
corresponding to helical motion around the z-axis.
Since the spinor equations are two coupled first order differential equations for the 2-spinors $\pi^A$ and $\eta^A$ and the Dirac equation is also first order involving a four-component spinor, it seems adequate to see if there is some connection between them. To this end we start with the expression of the four-momentum
\begin{equation} 
p^{AA'}=\frac{1}{\sqrt{2}}\left[\pi^{A}\bar\pi^{A'}+\eta^{A}\bar\eta^{A'}\right].
\end{equation} \\
Contraction of this equation with $\pi_A$ and of its covariant counterpart with $\bar\eta^{A'}$ leads to
\begin{equation}
\begin{array}{c}
p^{AA'}\pi_A = \frac{m}{\sqrt{2}}\overline\eta^{A'}\\
p_{AA'}\overline\eta^{A'} = -\frac{m}{\sqrt{2}}\pi_A, \\
\end{array}
\end{equation}\\
where use have been made of $\pi^A \eta_A = m,\ \bar\eta_{A'}\bar\pi^{A'} = -m$ which are consequence of (\ref{condition}).\\
Quantization follows from the transcription of $p^{AA'}$ to the hermitian operator:
\begin{equation}
\nabla^{AA'}
= \frac{1}{\sqrt 2}\left( \begin{array}{cc}
\partial_0+\partial_3 & \partial_1+i\partial_2 \\
\partial_1-i\partial_2 & \partial_0-\partial_3 \\
\end{array}\right),
\end{equation}\\
and reinterpretation of the spinors $\pi^A$ and $\eta^A$ and its complex conjugates as wave functions. The Dirac equation of a free particle in spinor form is thus \cite{penrose}
\begin{equation}
\begin{array}{c}
\nabla^{AA'}\pi _{A}=\frac {m}{\sqrt {2}}\overline\eta^{A'}\\
\nabla _{AA'}\overline {\eta }^{A'}=-\frac {m}{\sqrt {2}}\pi _{A}.\\
\end{array}
\end{equation}
The extension of the former equation to the corresponding of a particle interacting with an external electromagnetic field follows from the minimal coupling rule
\begin{equation}
\nabla^{AA'}\longrightarrow \nabla^{AA'}-eA^{AA'}.
\end{equation}\\
The coupling via the four-potential hermitian spinor $A^{AA'}$ manifest the essential difference with the spinor equations in which the action of the electromagnetic field is mediated by the $\vec E$ and $\vec B$ fields.

\section{$U(1)$ INVARIANT LAGRANGIAN}
Following the same line of development leading to the usual formulation of $U(1)$ gauge theory (based on 4-spinors and the Dirac Equation), it is convenient to define a lagrangian density, for a free particle, along the {\it classical} path of the particle (with dimension energy per unit length) as

\begin{equation} \label{lagrf}
{\cal L}=\dot{\eta}^{A}\pi_{A},
\end{equation} \\
together with the Euler Lagrange equations
\begin{equation} \label{1.3}
\frac{d}{d\tau}\frac{\partial \cal L}{\partial \dot{\eta}^{A}}-\frac{\partial \cal L}{\partial \eta^{A}}=0.
\end{equation} \\
With similar equations for the spinor $\pi^A$. The former equations lead to $\dot{\pi}_{A}=0 \Rightarrow \pi_{A}=const.$\\
We now consider the consequences of imposing invariance under local ({\it along the classical path parametrized by $\tau$}) phase transformations
\begin{equation} \label{1.6}
\begin{array}{c}
\eta^{A} \rightarrow e^{i\alpha(\tau)}\eta^{A} \\
\pi^{A} \rightarrow e^{i\xi(\tau)}\pi^{A}. \\
\end{array}
\end{equation} \\
The phase parameters $\alpha(\tau)$ and $\xi(\tau)$ cannot be independent as the spinors $\eta^A$ and $\pi^A$ are also not independent since from (\ref{momentum}), they are related by the condition \\
\begin{equation} \label{1.7}
p^{AA'}p_{AA'}=\vert\pi^{A}\eta_{A}\vert^{2}=m^{2}.
\end{equation}
{ } \\
In consequence $\eta^A\pi_A=const.$, leading to the constraint $\xi(\tau)=-\alpha(\tau)$.\\

As the classical trajectory should not be affected by any phase transformation, it is apparent that local gauge transformations leaves invariant the four-momentum of the particle:
\begin{displaymath}
p^{AA'}=\frac{1}{\sqrt{2}}\left[\pi^{A}\bar{\pi}^{A'}+\eta^{A}\bar{\eta}^{A'}\right].
\end{displaymath} \\
However, the free lagrangian (\ref{lagrf}) transform to \\
\begin{equation} \label{1.10}
{\cal L} \rightarrow i\dot{\alpha}\eta^{A}\pi_{A}+
\dot{\eta}^{A}\pi_{A}=i\dot{\alpha}\epsilon_{AB}\eta^B\pi^A+\dot{\eta}^A\pi_A.
\end{equation} \\
To find a gauge invariant lagrangian we have to add a term
\begin{equation} \label{1.12}
-\frac{e}{m}\phi_{AB}\eta^{B}\pi^{A},
\end{equation} \\
and impose the condition for the new field $\phi_{AB}$ of transforming, under local phase
transformations, as \cite{theorem}\\

\begin{equation} \label{1.13}
\phi_{AB} \rightarrow \phi_{AB}+i\frac{m}{e}\dot{\alpha}\epsilon_{AB},
\end{equation} \\
then, the new lagrangian \\
\begin{equation} \label{1.14}
{\cal L}=\dot{\eta}^{A}\pi_{A} -
\frac{e}{m}\phi_{AB}\eta^{B}\pi^{A},
\end{equation} \\
is invariant under $U(1)$ local-phase transformations. The transformation 
that holds
for the conjugate second-rank spinor   $\bar{\phi}_{A'B'}$,  is
given by \\
\begin{equation} \label{1.15}
\bar{\phi}_{A'B'} \rightarrow \bar{\phi}_{A'B'}-i\frac{m}{e} \dot\alpha\epsilon_{A'B'}.
\end{equation} \\
These kind of transformations leave however
invariant the associated
four-rank spinor of the Maxwell field strength \\
\begin{displaymath}
F_{ABA'B'}=\epsilon_{AB}\bar{\phi}_{A'B'}+\epsilon_{A'B'}\phi_{AB}.
\end{displaymath} \\
%
%
From the Euler Lagrange equations applied to the lagrangian given by (\ref{1.14}) it is immediate to obtain
\begin{equation}
\dot\pi_A=-\frac{e}{m}\phi_{AB}\pi^B.
\end{equation}
This equation and those given by (\ref{eqofmot}) are gauge invariant. This property will be used in next section for the specific case of the field created by a magnetic monopole.
\section{QUANTIZATION OF CHARGE}
The local gauge invariance of the equations of motion, through the transformation of $\phi^{AB}$, whose components are the electric and magnetic field strengths, can be applied to a long standing issue: The quantization of charge in the presence of a magnetic monopole  put forward by Dirac in 1931 \cite{dirac}. As it has always been  known, the root of the problem is the incompatibility of the vector potential definition $\vec B=\nabla\times \vec A$ with $\nabla.\vec B \ne 0$ when a magnetic monopole is the source of $\vec B$. To circumvent this problem Dirac \cite{dirac2} introduced later the notion of a string attached to the monopole. However, as was shown by Wu and Yang \cite{wu} in 1975, it is not possible to define a global singularity free $A^\mu$ in the presence of a monopole. The usual way out consist of defining two different four potentials $A^\mu$, in the north and south regions of the monopole (each singularity free in their respective regions), differing by a gauge transformation in the overlapping region and thus leading to the Dirac quantization condition ($2e\mu=n$, $\mu$ being the magnetic charge of the monopole), via the requirement imposed to the gauge transformation of being single-valued. The procedure just outlined and the hypothetical existence of the {\it strings} attached to the monopoles have been the subject of some controversy (see, for instance, the review by Preskill \cite{preskill}) which survives to the present day.

As already mentioned, the monopole vector potential cannot be smoothly defined everywhere in space. However, and since the electromagnetic field tensor $F^{\alpha\beta}$ is defined globally, perhaps the singularity of the vector potential can be avoided in an alternative mathematical description free of singularities of any kind. We shall show that this is indeed the case.

For a monopole of magnetic charge $\mu$ the field spinor $\phi^{AB}$, in spherical coordinates $(r,\theta,\varphi)$, from (\ref{phi}), is given by
\begin{equation}\label{monopole}
\phi^{AB} = i\frac{\mu}{2r^2}{\left(\begin{array}{cc} -\sin\theta e^{-i\varphi}& \cos\theta \\ \cos\theta & \sin\theta e^{i\varphi}\end{array}\right)}.
\end{equation} \\
Consider now one of the two equations of motion of an electron in the magnetic field of the monopole, for instance
\begin{equation}\label{equ}
\dot\eta^A=\frac{e}{m}\phi^{AB}\eta_B.
\end{equation} \\
As we are free to choose the $U(1)$ group parameter, we take the azimuthal angle $\varphi$, and consider the usual gauge transformation (found in many text books and elsewhere):
\begin{equation}
\eta^A \rightarrow \eta^A e^{ie\varphi}.
\end{equation} \\
Invariance of (\ref{equ})  requires a transformation of $\phi^{AB}$ according to
\begin{equation}
\phi^{AB} \rightarrow \phi^{AB}+i m \dot\varphi\ \epsilon^{AB}
\end{equation}
However, since the electric charge $e$ is not an integer (in almost every system of units), a phase factor of the form $exp{[ie\varphi]}$ produce an inconsistent result when we perform a $2\pi$ rotation (for fixed $r$ and $\theta$) $\varphi \rightarrow (\varphi+2\pi)$, as $\phi^{AB}$, given by (\ref{monopole}), return to its original value while the phase factor $exp[ie\varphi]$ does not. We are thus lead to consider only irreducible representations of the group $U(1)$, isomorphic to the compact \cite{compact} circle group {\bf T}, which, as is well known, are of the form
\begin{equation}\label{c.c.}
e^{in\varphi}
\end{equation}\\
with $n \in Z$.

In view of the precedent considerations, in the presence of a monopole, the natural prescription for a quantization rule would be $e\mu=n$, which was proposed by Schwinger \cite{schwinger} some years ago and correspond to the dimensionless coupling constant $e\mu$ ($e\mu /\hbar c$ in conventional gaussian units) between the electron and the monopole. However, this condition, as it stands, is inconclusive since a gauge transformation $exp{(i2e\mu\varphi)}$ also leaves (\ref{equ}) invariant and leads to the quantization condition of Dirac $2e\mu=n$.

 To proceed further let us locate the electron at rest either in the positive or negative z-axis ($\theta=0$ or $\theta=\pi$) at some distance $r$. Solving the equations of motion for both spinors $\eta^A$ and $\pi^A$, with the constraint condition $\eta^A\pi_A=m$, we find (in qualitative agreement with the solution of the Dirac equation for an electron at rest in a constant magnetic field \cite{bjorken})

\begin{equation}\label{sol1}
\pi^{A}(\tau)=\sqrt{m}\left(\begin{array}{c}
e^{-i\frac{e\mu}{2mr^2}\tau} \\
0 \\
\end{array}\right),\quad \eta^{A}(\tau)=\sqrt{m}\left(\begin{array}{c}
0 \\
e^{i\frac{e\mu}{2mr^2}\tau} \\
\end{array}\right),
\end{equation} \\
for the positive axis, the solutions for the negative z-axis being

\begin{equation}\label{sol2}
\pi^{A}(\tau)=\sqrt{m}\left(\begin{array}{c}
e^{i\frac{e\mu}{2mr^2}\tau} \\
0 \\
\end{array}\right),\quad \eta^{A}(\tau)=\sqrt{m}\left(\begin{array}{c}
0 \\
e^{-i\frac{e\mu}{2mr^2}\tau} \\
\end{array}\right).
\end{equation} \\
As the choice of the positive (negative) axis is a matter of comvenience, reflected in the Maxwell spinor $F^{ABA'B'}$ being the same in both situations (in tensor lenguage $F^{\alpha\beta}$ is not). If we take one of the spinors, $\eta^A$ for instance, we see that both solutions are related by the gauge transformation

\begin{equation}
\eta^A \rightarrow \eta^A e^{-i \frac{e\mu}{mr^2}\tau}
\end{equation} \\
Together with

\begin{equation}
(\phi^{AB})^{+}= (\phi^{AB})^{-} - \frac{m}{e}\left( \frac{d\Omega}{d\tau}\right) \Omega^{-1},
\end{equation} \\
with $(\phi^{AB})^{+}$ and $(\phi^{AB})^{-}$ corresponding to the positive and negative z-axis respectively and both related by the gauge transformation

\begin{equation}
\Omega=e^{-i\frac{e\mu}{mr^2}\tau}.
\end{equation} \\

Let us now examine under what conditions $\Omega$ can be written in the form given by (\ref{c.c.}) satisfying the irreducible representation condition of the compact group $U(1)$.
First we note that $e\mu/mr^2$ is the Larmor frequency $\omega_L$ (with a g-factor 2) due to the spin precession at the electron location. If the Schwinger quantization condition $e\mu=n$ 
($e\mu /\hbar c=n$ in comventional Gaussian units) holds, for $n=1$:

\begin{equation}
\Omega=e^{-ie\mu\omega_L\tau}=e^{-i\omega_L\tau}=e^{-i\varphi},
\end{equation} \\
and for any integer $n$

\begin{equation}
\Omega=e^{-i n\omega_L\tau}=e^{-in\varphi}.
\end{equation}\\
 The integer $n$, the magnetic (electric) charge in units of $1/e$ ($1/\mu$), is a winding number: the number of times $\Omega(\varphi)$ covers the $U(1)$ gauge group as $\varphi$ varies from 0 to $2\pi$. In this picture, the electric and magnetic coupling constants are dual to each other.


Upon substitution of the Larmor frequency $\omega_L$ in the spinor solutions of the equations of motion it is apparent the spin one-half behaviour of the electron of returning to itself after a $4\pi$ rotation.

To end a few comments about the Dirac condition. Curiously enough, Dirac in his first and subsequents studies about the subject did not take into consideration the electron spin. Other approaches  also ignored the possibility of a certain dependence with  the spin (There is an early, nonrelativistic, contribution of Goldhaber \cite{goldhaber} to the role of spin in the monopole problem). Although the spinor equations are only valid for spin one-half particles, we can tentatively and in an heuristic way argue that for a charged particle with intrinsic spin 1 (a $W^+$ for instance) the solutions of the equations of motion, being classical and equivalent to the Lorentz force albeit incorporing internal degrees of freedom, could be used (Note that the group $SL(2,C)$ include representations of half-integer spins and integer spins \cite{carmeli}). However, for a spin 1 particle  the Larmor frequency (g-factor=1) is
\begin{equation}
\omega_L=\frac{e\mu}{2mr^2}.
\end{equation}
The gauge transformation would be then
\begin{equation}
\Omega=e^{-i2e\mu \omega_L\tau}=e^{-i2e\mu\varphi},
\end{equation}\\
leading to the Dirac quantization condition.

\end{document}